\documentclass[doublecol]{epl2} 
\usepackage{graphicx}
\usepackage{amsmath}
\usepackage{amssymb}
\usepackage{bm}     
\usepackage{subfigure,pslatex}
\usepackage{mathrsfs} % for scripted letters
\usepackage{color}
\title{Fluctuations in Shear-Jammed States: A Statistical Ensemble Approach}
\shorttitle{Fluctuations in Shear-Jammed States: A Statistical Ensemble Approach} %Insert here a short version of the title if it exceeds 70 characters

\author{Dapeng Bi \inst{1} \and Jie Zhang \inst{2} \and R. P. Behringer \inst{3} \and Bulbul Chakraborty \inst{4}}
\shortauthor{D. Bi \etal}

\institute{                    
  \inst{1} Department of Physics, Syracuse University, Syracuse, New York 13244 USA\\
  \inst{2} Institute of Natural Sciences and Department of Physics, Shanghai Jiao Tong University, Shanghai 200240 China\\
  \inst{3} Department of Physics, Duke University, Durham, North Carolina 27708, USA\\
  \inst{4} Martin Fisher School of Physics, Brandeis University, Waltham, Massachusetts 02454 USA
}

\abstract{
Granular matter exists out of thermal equilibrium, i.e. it is athermal. While conventional equilibrium statistical mechanics is not useful for characterizing granular materials, the idea of constructing a statistical ensemble analogous to its equilibrium counterpart to describe static granular matter was proposed by Edwards and Oakshott more than two decades ago. Recent years have seen several implementations of this idea. One of these is the stress ensemble, which is based on properties of the force moment tensor, and applies to frictional and frictionless grains. We demonstrate the full utility of this statistical framework in shear jammed (SJ) experimental states\cite{Nature_Bi,Behringer_Dijksman_Ren_PRL}, a special class of granular solids created by pure shear, which is a strictly non-equilbrium protocol for creating solids. We demonstrate that the stress ensemble provides an excellent quantitative description of fluctuations in experimental SJ states. We show that the stress fluctuations are controlled by a single tensorial quantity: the angoricity of the system, which is a direct analog of the thermodynamic temperature. SJ states exhibit significant correlations in local stresses and are thus inherently different from density-driven, isotropically jammed (IJ) states. 
}

\begin{document}
\maketitle

\subsection{Introduction}
A remarkable property of systems  in thermal equilibrium is that the probability of occurrence of a microscopic state is known {\it a priori} through the universal Boltzmann distribution: $P_{\nu} = e^{-\beta E_{\nu}}/Z(\beta)$.  Here, $\beta$ is the inverse temperature and $E_{\nu}$ is the energy of microstate $\nu$.  The Boltzmann distributions defines characteristics such as the relation between fluctuations and response\cite{Parisi_book}. By contrast, granular systems are intrinsically out of thermal equilibrium, and we lack a broad framework for describing their statistical properties\cite{Bouchaud_lecture}.  The idea of constructing statistical ensembles to describe granular systems originated in a proposal by Edwards\cite{Edwards_Oakeshott} that the fluctuations of slowly-driven, dense
granular systems was controlled by the ensemble of ``blocked'' states,  granular assemblies in static, mechanical equilibrium.  The original Edwards ensemble uses free volume in a granular system as the analog of energy in a Gibbsian statistical framework.  
A more recent model is the force network ensemble (FNE) \cite{Snoeijer2004,tighe}, which has been useful in describing force fluctuations on fixed granular geometries.  The stress ensemble is a generalization of the original Edwards idea, and is based on a ``conservation'' principle that arises from the constraint of {\it local} force and torque balance on every grain\cite{BC_softmatter}.  Merging of the original Edwards ideas with the stress ensemble approach leads to an analog of the Boltzmann distribution for granular solids (blocked or jammed states), where the role of temperature is played by two distinct quantities: (a) the compactivity which is conjugate to volume,  and (b) angoricity, a tensor that is conjugate to the force-moment tensor of granular solids\cite{Henkes2009,Blumenfeld_PRL2012}.  

The applicability of the concept of angoricity as a temperature-like variable has been beautifully demonstrated in recent experiments using photoelastic disks\cite{Daniels-Puckett}.  In earlier work, the concept of angoricity was applied to analyze stress fluctuations in simulated assemblies of frictionless grains\cite{Henkes2009,Henkes2007}.   These granular solids,  frictional assemblies in experiments, and frictionless in the simulations, shared the common feature that the solidification is density-driven.  They, therefore fall within the rubric of the universal jamming phase diagram\cite{LiuNagel}, and can be viewed as zero-temperature limits of thermal, amorphous solids.   

A class of granular solids that are far-from-equilibrium, and not described within the original jamming scenario are shear-jammed (SJ) states.\cite{Nature_Bi,Behringer_Dijksman_Ren_PRL}.   These solids are created through shearing without changing the density, which is a strictly non-equilibrium pathway for creating solids.      They provide a unique opportunity for testing the stress ensemble with its full tensorial complexity in an ensemble that has no equilibrium analog.    In this work we show that the stress fluctuations in SJ states are described quantitatively by the stress ensemble, which is the infinite compactivity limit of the generalized ensemble.   This test of the stress ensemble is non-trivial since we demonstrate that stresses have non-trivial correlations in SJ states, unlike in density-driven granular solids.

%In jammed granular solids, the dissipative and athermal nature prevents any statistical descriptions that are based on energy conservation. By adopting a well defined experimental or numerical protocols, an ensemble of grain configurations can be produced that are static and mechanically stable. It has been shown that with a careful choice of a protocol \cite{Nowak_shaking,corey_simul}, the ensemble of states formed is reproducible \cite{FNE,Nowak_shaking,corey_simul}. Even in the absence of dynamics, this suggests the possibility of using a statistical ensemble approach to analyze the fluctuation in static and mechanically stable states (fluctuations here refer to sample-to-sample and to spatial fluctuations within one realization). Over the years, many approaches have been proposed, for example the earliest is the Edwards ensemble, which uses free volume in a granular system as the analog of energy in a Gibbsian statistical framework. Another model is the force network ensemble (FNE) \cite{Snoeijer2004,tight}, which has been useful in describing force fluctuations on fixed granular geometries. In this paper, we will focus on the stress ensemble framework that is a generalization of the Edwards idea, and based on the existence of a conserved quantity, the force moment tensor \cite{ball-blumenfeld, Henkes}. Specifically, we extend the application of the stress ensemble to experimental shear jammed states \cite{Nature_bi}. 

The paper is organized as follows:  Section 2 provides a brief review of experimental SJ states, Section 3 introduces the tensorial formulation of the stress ensemble framework, Section 4 presents results of testing the stress ensemble framework in SJ states, and finally, in Section 5, we discuss the correlations in SJ states, and its implications.

\subsection{Shear Jammed (SJ) states}
Dry grains interact with purely repulsive, contact  interactions.  Without any cohesion between them, the only way to create a solid of dry grains is by applying a load at the boundary. 
%Traditionally jammed states are created  using either numerical (e.g. see \cite{Trush_isotropic}) or numerical (e.g. \cite{corey_simul}) techniques. 
A common feature of  most experimental \cite{Daniels-Puckett, Trush_isotropic} and numerical \cite{corey_simul} techniques for creating jammed states is the application of isotropic pressure (we shall call IJ states from here on). The jamming transition is defined as the onset of mechanical stability, and usually associated with a particular packing fraction, $\phi_J,$\cite{LiuNagel} that can be protocol dependent. IJ states can also be sheared  to study their response \cite{corey_simul,Olsson:2007p243}. However, in these cases, shear stress should be considered a perturbation to an already jammed state, and under large enough shear stress, IJ states can unjam into flowing states.

%%%
\begin{figure}[htbp]
\begin{center}
\includegraphics[width=0.49\textwidth]{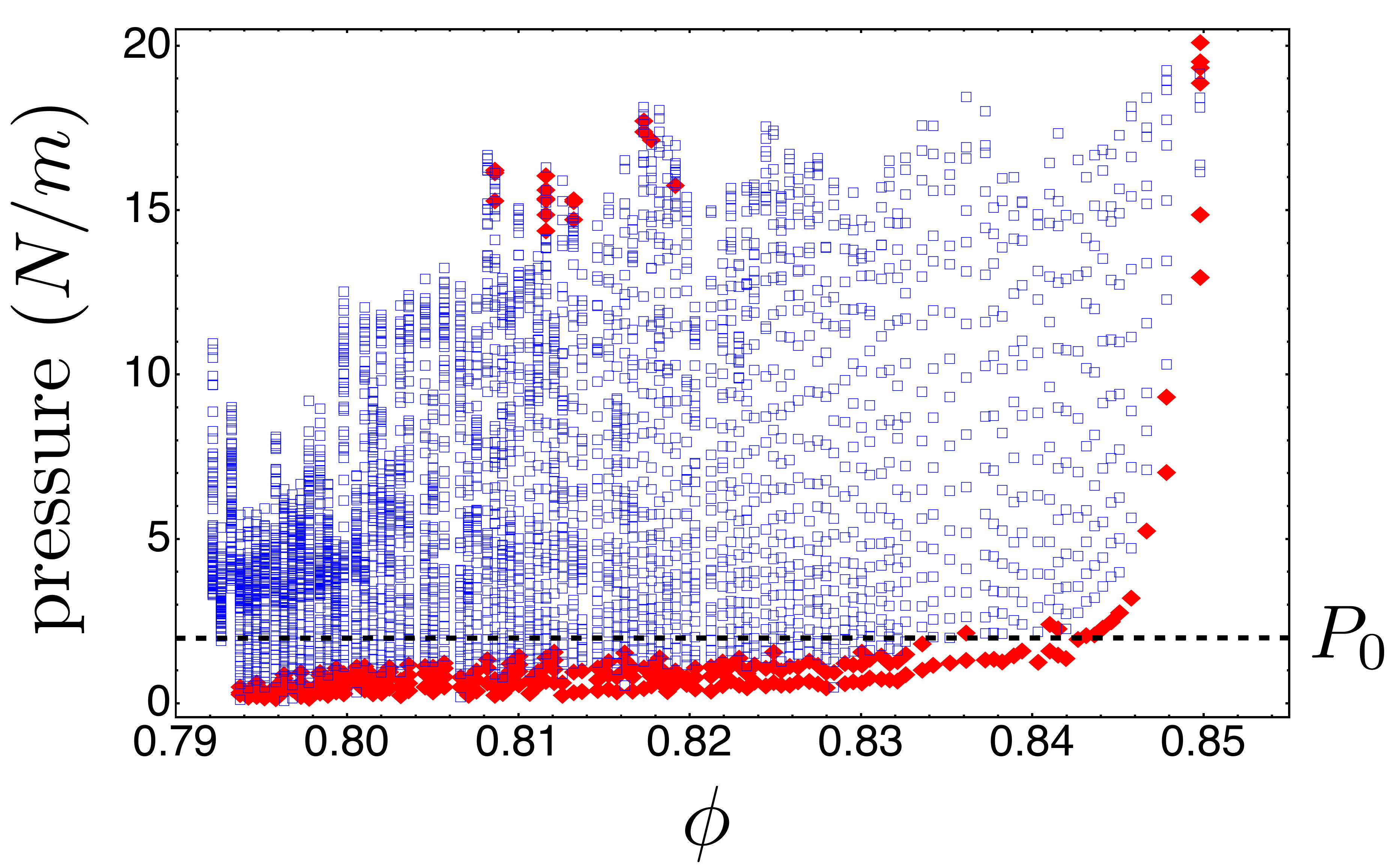}
\caption
{
Plot of the Pressure versus the packing fraction for all experimental states. Red: isotropic states with $\tau \approx 0$. Blue: states with $\tau > 0$. 
The experimental resolution for the pressure is $P_0$. below this threshold, there is uncertainty to weather a state actually has finite pressure, and is in fact jammed,  or if the pressure is caused by friction with base plate. However, what is {\it certain} is that there are no isotropic states below $\phi_J$, for these states there is a sharp transition to finite pressure at $\phi_J$. For states with shear stress, there is no sharp transition as a function of the packing fraction.
}
\label{p_v_phi}
\end{center}
\end{figure}
%%%
In contrast, the observation of SJ states in experiments \cite{Nature_Bi,Behringer_Dijksman_Ren_PRL} and simulations \cite{luding-sj} has demonstrated that applying isotropic pressure is not a necessary condition for jamming. For a range of packing fractions below the isotropic $\phi_J$ for a given protocol, jammed states can be created by applying shear only. The experimental SJ states, in two dimensions,  are created through application of quasistatic, forward shear\cite{Nature_Bi}.
%These states created  are more anisotropic in their force and contact networks than the IJ states \cite{Nature_Bi}. 
Packing fraction does not play a significant role in determining the properties of SJ states. For example, it has been shown that pressure and shear stress do not depend monotonically on $\phi$ as one would expect in IJ states, but rather monotonically scale with the fraction of force bearing grains in the system\cite{Nature_Bi}. The SJ states are qualitatively different from IJ states.  Whereas IJ states can have non-vanishing pressure and zero shear stress, pressure arises as a consequence of shearing in SJ states\cite{Behringer_Dijksman_Ren_PRL}.  SJ states have strongly anisotropic fabric and force networks and stress tensors. It has been shown that the SJ states emerge as a consequence of the percolation transition in the force network\cite{Nature_Bi}.
%These properties make SJ states and an ideal candidate for testing the broad predictions of the theoretical framework of the stress ensemble with its full tensorial complexity.   Applications of the stress ensemble to IJ states created by simulations \cite{Henkes}, and experiments\cite{James and Karen}, have successfully tested the concept of an intensive temperature-like quantity, the angoricity, but only in its scalar form.   The SJ states can be grouped together by their full stress tensor to form an ensemble of states with the force-moment tensor, and measured fluctuations can be compared to the predictions of the stress ensemble..

\subsection{Stress Ensemble}
Over a decade ago, Edwards and Oakeshott \cite{Edwards_Oakeshott} theorized that the dynamics of slowly driven granular materials is controlled by the statistics of mechanically stable configurations, know as blocked states. A simple thermal system is specified by the state variables $E$,$V$, and $N$.  
In constructing a statistical ensemble for athermal granular, the role of the Hamiltonian in a thermal system is replaced by the volume function of a granular packing. Known as the Edwards ensemble, all blocked states of the same volume $V$ are assumed to be equiprobable (microcanonical hypothesis). An entropy is defined for a granular state with the state variables $V$ and number of grains $N$, $S(V,N)$. The analog of the thermodynamic temperature can be defined by $X(V)=\partial V / \partial S$. $X$ or the compactivity \cite{Edwards_Oakeshott} is conjugate  to the volume, and hence $X=X(V)$ is also the primary equation of state. A microstate labelled by $\nu$, inside a granular packing with total volume $V$ is found with a probability proportional to the canonical distribution
\begin{equation}
P_{\nu} =\frac{1}{Z(X)}  \exp [-w_{\nu} / X].
\label{edwards_canonical}
\end{equation}
Here, the  volume function (analog of the energy $E_{\nu}$ s determined by the set of grain positions $w_\nu=w(\{ \vec{r}_i \})$.

Recent interest has focused on
conservation laws and the sampling of
the phase space of blocked states, especially in the context of extending the
ensemble framework to stiff but not infinitely rigid grains. The
basic aim of any ensemble approach is to predict the probability of
occurrence of a microscopic state, given a set of macroscopic,
measurable quantities such as the volume and the external stress.
A static granular packing is ultimately characterized by the positions of the
grains, $\{ \vec{r}_i \}$,  and the set of contact forces, $\{ \vec{f}_{ij} \}$.  Specification of the forces is essential  for defining static states of frictional grains, and  it is therefore crucial to include the contact forces in a {\it generalized} statistical description for infinitely rigid and stiff grains. Among the various formulations there are two different ways to incorporate $\{ \vec{f}_{ij} \}$ in an ensemble, (1) the FNE \cite{Snoeijer2004} in which micro states are characterized only by $\{ \vec{f}_{ij} \}$, and (2) the stress ensemble \cite{BC_softmatter, Henkes2009, Blumenfeld_PRL2012} \footnote{We emphasize that the stress ensemble described here is not on a fixed geometry, but rather comprises of states with different contact and force networks.}, which considers $\{ \vec{r}_i \}$,  and $\{ \vec{f}_{ij} \}$. The latter, unlike the FNE, is not restricted to a fixed geometry

In the stress ensemble, in addition to volume, granular blocked states are also characterized by the force-moment tensor, which is a function of both grain positions $\{ \vec{r}_{ij} \}$ and inter-granular forces $\{ \vec{f}_{ij} \}$
\begin{equation}
\hat{\Sigma}=
\sum_{i<j} \vec{r}_{ij} \otimes \vec{f}_{ij}
\overset{2d}{=}
\begin{pmatrix}
\Sigma_{11} & \Sigma_{12}  \\
\Sigma_{12} & \Sigma_{22}
\end{pmatrix}
.
\label{force-moment}
\end{equation}
The constraint of local force balance leads to a  conservation law for $\hat \Sigma$\cite{BC_softmatter,Daniels-Puckett}, which states that the force-moment tensor cannot be changed by local intervention such as tapping.  Changing $\hat \Sigma$, while maintaining force balance on every grain, requires changing the forces on a line of grains that spans the system\cite{BC_softmatter,Henkes2009}. For a large system, there is thus a conservation principle reminiscent of energy conservation in non-dissipative systems.  Taking tapping as an example for generating blocked states, the microcanonical stress ensemble is defined by fixing the values of $V$ and $\hat{\Sigma}$.  
In addition to compactivity, the stress ensemble involves an intensive variable that is conjugate to $\hat{\Sigma}$.  This is the angoricity tensor, $\hat{\alpha}^{-1}$ identified by Edwards and coworkers \cite{Edwards_Oakeshott}. The relation between $\hat \alpha$ and $\hat \Sigma$ is analogous to the relationship between $\beta$ and $E$ in a thermodynamic system. 

In a canonical ensemble formulation\cite{Henkes2009},  the probability of obtaining a microstate $\nu$ with volume $w_\nu$ and force moment tensor $\hat{\Sigma}_\nu$ is given by a {\it tensorial} generalization of the Boltzmann distribution
\begin{equation}
P_{\nu}=\frac{1}{Z(X,\hat{\alpha})} ~ \exp \left(-\frac{w_{\nu}} {X}\right) ~ \exp \left(-\hat{\alpha}:\hat{\Sigma}_{\nu}\right),
\label{generalized_ensemble}
\end{equation}
where ``:'' means tensor contraction and the angoricity tensor $\hat{\alpha}^{-1}$ is defined by
\begin{equation}
\alpha_{k l}=\frac{\partial{S(V, \hat{\Sigma})}}{\partial{\Sigma_{k l}}} \bigg|_{V} .
\label{angoricity_definition}
\end{equation}
In two dimensions,  we have 
\begin{equation}
\hat{\alpha}=
\begin{pmatrix}
\alpha_{11} & \alpha_{12}  \\
\alpha_{12} & \alpha_{22}
\end{pmatrix}.
\label{angoricity_2d}
\end{equation}
In the above thermodynamic formulation, the entropy $S(V, \hat{\Sigma})$ should be taken as the large $M$ limit of the entropy of $M$ grains.  
Since all blocked states are in force and torque balance, their force moment tensor $\hat{\Sigma}$ is always symmetric, hence $\hat{\alpha}$ is also symmetric via eq.~\ref{angoricity_definition}. It should also be pointed out that while both $\hat{\alpha}$ and $\hat{\Sigma}$ transform like tensors under rotation, the argument of the Boltzmann term in eq.~\ref{generalized_ensemble} contains only terms that are rotationally invariant. 

%%%
\begin{figure}[htbp]
\begin{center}
\includegraphics[width=0.49\textwidth]{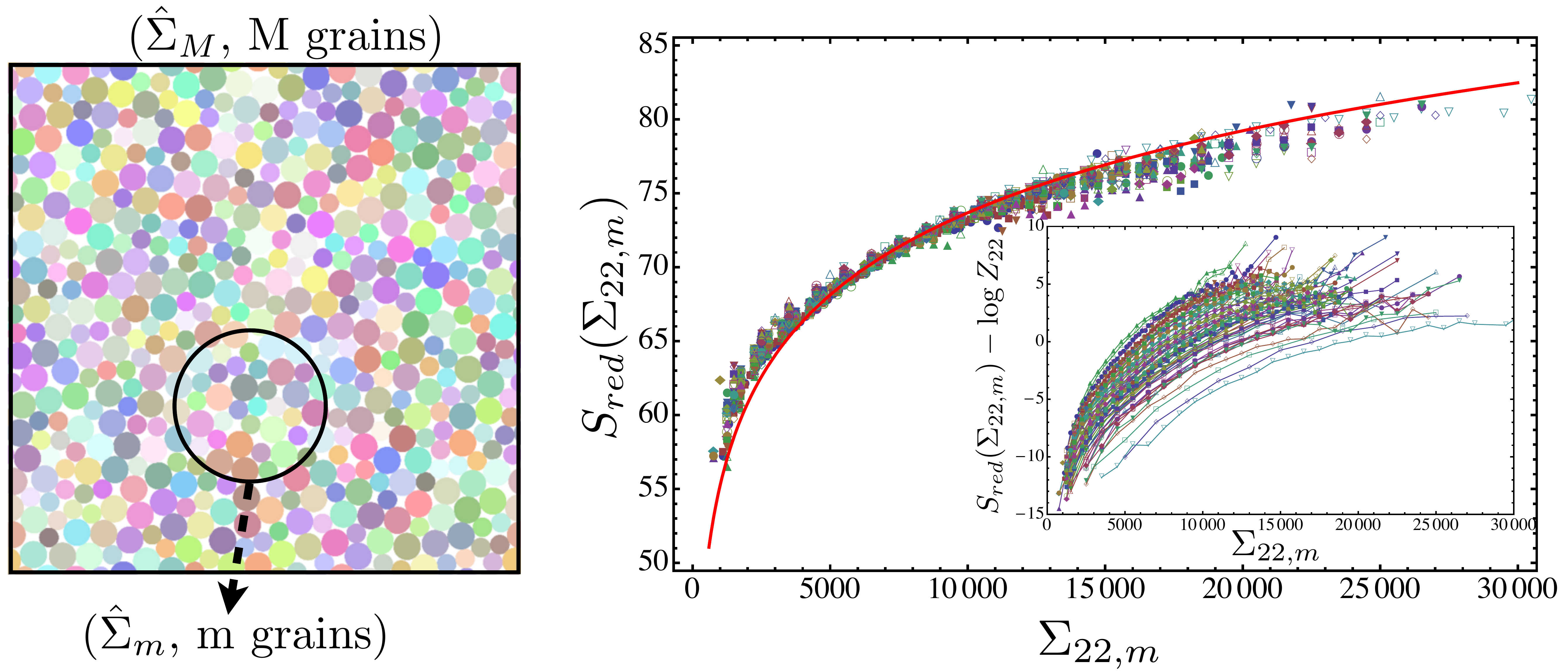}
\caption
{
(a)
Schematic demonstration of a sub-region of size $m$ and  $\hat{\Sigma}_m$ inside a packing with ($M$, $\hat{\Sigma}_M$).
Here we define subregions as any contiguous cluster of $m$-grains roughly circular in shape.
%(b) Inset: for $m=15$ the distribution of the $11-$component of the local force moment tensor, $\sigma_{11,m}$ is plotted for various values of $\Sigma_{11,M}$. With proper rescaling (eq.~\ref{rescaled_local_fmt}) of the local force moment by the angoricity, the distributions collapse onto one single curve. 
(b) Eq.~\ref{full_canonical} predicts that when the distribution of local force moment tensor is multiplied by $e^{\hat \alpha : \hat \Sigma}$ (shown in inset for $m=15$ and the $22-$component of the local force moment tensor in SJ states), its logarithm should be related to the entropy of microstates $S_m$, and for different $\hat{\alpha}'s$ this function should only differ by an additive constant. 
%In the inset, we show the unshifted distributions  for $m=15$ and the $22-$component of the local force moment tensor in SJ states. 
After a shift of each curve by $Z_{22}$ calculated from the eq.~\ref{reduced_xxyy_distributions}, we find the data collapse onto a universal curve in the main figure. The functional form of the collapses data also shows good agreement with the postulated form of entropy (eq.~\ref{reduced_entropy}), given by the solid red line.  
}
\label{combo_subregion_entropy}
\end{center}
\end{figure}
%%%

%An impotant point to mention is that Z(X,\alpha) cannot, in general, be written as a product Z(X) Z(\alpha) since \hat \Sigma depends on the grain positions\cite{recent_blumenfeld}.
Volume fluctuations in SJ states are less relevant than they are in IJ  states.  Analysis of SJ states shows that stress fluctuations and volume fluctuations are decoupled, and unlike IJ states, the packing fraction is not a state variable that determines the jamming probability under the shearing protocol\cite{Nature_Bi}.  These observations suggest that the pure stress ensemble, obtained in the limit of $X \to \infty$ is the one relevant for describing the fluctuations of SJ states. For a subset $m < M$ of the particles, the pure stress ensemble \cite{Henkes2009} is defined by the canonical distribution 
%{\color{red} changed all $\Omega$ to $S$, no $\Omega$ will appear through out the paper, is this okay?}
\begin{equation}
P(\hat{\Sigma}_m)=\frac{1}{Z(\hat{\alpha})} 
			     ~\exp{\left[S_m(\hat{\Sigma}_m) \right]}
			     ~ \exp{\left(-\hat{\alpha}:\hat{\Sigma}_m \right)}.
\label{full_canonical}
\end{equation}
The canonical distribution (eq.~\ref{full_canonical}) describes the stress fluctuation in a region containing the $m$ grains that is in a ``angoricity" bath due to the remainder of the system (see fig.~\ref{combo_subregion_entropy}(a)). $S_m(\hat{\Sigma}_m)$ is the (microcanonical) entropy function which is defined by counting the number of microstates 
\begin{equation}
S_m(\hat{\Sigma}_m) \propto \log \left [\sum_{\nu} ~ \omega_{\nu}~ \delta(\hat{\Sigma}_m-\hat{\Sigma}_{\nu}) \right].
\label{entropy_def}
\end{equation}
In this formulation, we have allowed for the possibility of microstates not being sampled equally by including a microscopic weight $\omega_\nu$ for state $\nu$ \cite{dauchot_bertin}.  The stress ensemble and the original Edwards ensemble can be extended to include this possibility\cite{mcnamara_volume,BC_softmatter}.

%\begin{equation}
%P(\hat{\Sigma}_m)=\frac{1}{Z(\hat{\alpha})} 
%			     ~\Omega(\hat{\Sigma}_m)
%			     ~ exp \left(-\hat{\alpha}:\hat{\Sigma}_m \right)
%\label{full_canonical}
%\end{equation}
%where $m$ will be used throughout to indicate the number of grains in the local subregion and 
%$\Omega(\hat{\Sigma}_m)$ is the microcanonical partition function and $\omega_\nu$ the microscopic weight for state $\nu$
%\begin{equation}
%\Omega(\hat{\Sigma}_m)=\sum_{\nu} ~ \omega_{\nu}~ \delta(\hat{\Sigma}_m-\hat{\Sigma}_{\nu}).
%\end{equation}
%In this formulation, we have allowed for the possibility of microstates not being sample equally\cite{dauchot_bertin}.  The stress ensemble and the original Edwards ensemble can be extended to include this possibility\cite{mcnamara_volume,BC-softmatter}.

\subsection{Measuring Angoricity}
In order to apply eq.~\ref{full_canonical} to the experimentally generated SJ states, we need a measure of $\hat \alpha$.   Unfortunately, we have no `thermometer' for directly measuring angoricity, and have to base its measurement on postulated or measured equations of state relating $\hat \alpha$ to $\hat \Sigma$.  

In IJ states, a scalar version of eq.~\ref{full_canonical} is applicable, and there is a scalar  $\alpha$ that is conjugate to $\Gamma$,  the trace of the force-moment tensor.
In \cite{Henkes2007}, the stress ensemble was applied to analyze pressure fluctuations in simulated frictionless IJ states, and an equation of state was deduced from these measurements.  More recently, an equation of state has been deduced from measurements in IJ states of frictional packings\cite{Daniels-Puckett}.   In both systems, a linear relation was found between the inverse angoricity and $\Gamma_M$ of an $M$ grain system:
%A simple scalar version of the canonical distribution (eq.~\ref{full_canonical}) was tested
%\begin{equation}
%P(\Gamma_m)=\frac{1}{Z(\alpha)} 
			     %~exp{\left[S_m(\Gamma_m) \right]}
			     %~ exp [-\alpha \Gamma_m],
%\label{Gamma_canonical}
%\end{equation}
%where only the trace of $\Gamma_m=tr(\hat{\Sigma}_m)/2$ was used. 
%An equation of state relating the inverse angoricity to the extensive $\Gamma_M$ of the packing was shown to be 
%\begin{equation}
$\alpha \approx \frac{ M}{\Gamma_M}$.
Since this linear relation worked quite well for pressure fluctuation in IJ states, we take a straightforward  generalization of this as a postulate for the equation of state for SJ states. 
%
%
%Since the ideal-gas hypothesis works quite well for pressure fluctuations in SJ states, we make an assume this also holds true for full tensorial case of the stress ensemble, we next make an educated guess for the equation of state based on Eqs.~\ref{silke_eos}: 
\begin{equation}
\hat{\alpha}
\propto
M \left(\hat{\Sigma}_{M}\right)^{-1}
=\frac{M}{\det\left (\hat{\Sigma}_M \right)}
		\begin{pmatrix}
			\Sigma_{{22},M} & -\Sigma_{{12},M}  \\
			-\Sigma_{{12},M} & \Sigma_{{11},M}
		\end{pmatrix}.
\label{EOS}
\end{equation}
Using this equation of state, each $M$-grain SJ state can be labelled by an angoricity tensor, since the force-moment tensor is measured in the experiments.  In the experiments, $M\sim1000$, and one could wonder about this being large enough to be in the thermodynamic limit.  We will address this question below through analysis of the convergence of the entropy function.

Eq.~\ref{full_canonical} predicts the distribution of stress in a subregion containing  $m$ grains at the angoricity given by $\hat{\alpha}$.    As has been pointed out before\cite{mcnamara_volume}, this Boltzmann-like distribution has a very special form.  It has a term that depends purely on $\hat \Sigma$, a term that depends purely on $\hat \alpha$, and the combination of these two variables appears only in the exponential.  The stress ensemble would, therefore, predict that multiplying the distribution by $e^{\hat \alpha : \hat \Sigma}$  and taking its logarithm would yield a function that depends on $\hat \Sigma$ and distributions corresponding to different $\hat \alpha$s would differ only by an additive constant, or 
\begin{equation}
\log\left[ P(\hat{\Sigma}_m)  e^{\hat{\alpha} : \hat
{\Sigma}_m} \right] = S_m\left(\hat{\Sigma}_m\right)-\log Z(\hat{\alpha})
\label{entropy_eq}
\end{equation}

We directly test these predictions as follows. All SJ states, regardless of the value of the shear strain or the packing fraction $\phi$, are categorized by their global force-moment-tensor $\hat{\Sigma}_M$. Through the equation of state (eqs.~\ref{EOS}), the value of the angoricity tensor is determined. Then for each SJ state at a particular $\hat{\alpha}$, random contiguous clusters containing $m$ grains (fig.~\ref{combo_subregion_entropy}(a)) are chosen to form an ensemble of subregions with different values of $\hat{\Sigma}_m$,  giving a probability distribution $P_{\hat{\alpha}} \left( \hat{\Sigma}_m \right)$. To simplify the analysis, we avoid dealing with the multi-dimensional distribution function
%, $P_{\hat{\alpha}} \left( \hat{\Sigma}_m \right)$, 
and instead, we analyze the reduced distributions for each component of the local force moment tensor. 

Fig. \ref{combo_subregion_entropy}(b) shows that the data collapse implied by eq.~\ref{entropy_eq} works remarkably well for a subregion containing $15$ grains.  We find similar data collapse for the other components of the force moment tensor, and for $m$ as small as $4$. 
%{\bf Max: check statements here}. This is correct. 
For smaller $m$, there is more of a spread in the data, however, there is remarkably fast convergence as a function of $m$ to a universal functional form, which we can interpret as the thermodynamic entropy $S(\Sigma_{kl})$ for $kl = 11, 22, 12$.  The solid red line in fig.~ \ref{combo_subregion_entropy}(b) is the functional form deduced from the assumed equation of state and the definition of  of $\hat{\alpha}$ (eq.~\ref{angoricity_definition})
\begin{equation}
S_M(\hat{\Sigma}_M) = A \ M \ \log \left[ \det \ \hat{\Sigma}_M \right].
\label{generalized_entropy}
\end{equation}
Here, $A=1/2$ is a constant of proportionality that is determined from the best fit to the collapsed data.  A few features of fig.~ \ref{combo_subregion_entropy}(b) are of note:  there is a small but systematic difference between the entropy deduced from the equation of state, and the form of the collapsed data, especially at small values of $\Sigma_{kl}$.   In spite of these small differences, the fact that we can collapse the data demonstrates that the Boltzmann-like distribution defined by an angoricity tensor (eq.~\ref{full_canonical}) works remarkably well for SJ states.

The equation of state that we postulated is the analog of an ideal-gas equation of state relating temperature and energy.   If the entropy implied by this equation of state (eq.~\ref{generalized_entropy}) were to hold exactly for the SJ states, then it would imply that stress correlations in these states are shorter-ranged than the size of the $m$-grain subregions that we have analyzed.  Below, we assume that the ideal-gas entropy holds, deduce an explicit scaling form for the distribution of stresses for $m$-grain subregions, and show that the ideal gas model fails  to describe the distributions in detail, and especially their scaling with $m$.    To understand this observation, we analyze the variance of the components of the stress, and what it implies for stress correlations, and finally we discuss how to reconcile the success of the equation of state with these observations.

Since the SJ states are created with bi-axial shear the resulting normal stress $\Sigma_{11,M}-\Sigma_{22,M}$ is much larger ($\sim 10$ times \cite{Nature_Bi}) compared to the off-diagonal component $\Sigma_{12,M}$.  This justifies expanding the equation of state, eq.~\ref{EOS} to first order in
\begin{equation}
\frac{{\left(\Sigma_{12,M}\right)}^2} {\Sigma_{11,M} \ \Sigma_{22,M}} \ll 1. 
\label{small_xy_limit}
\end{equation}
Then, reduced distributions of a particular force moment tensor component $\Sigma_{kl}$ can be obtained from eq.~\ref{full_canonical} and eq.~\ref{generalized_entropy} by integrating out all other components $mn \ne kl$. For the $11$ and $22$ components
 \begin{equation}
	\begin{split}
	p_{11,m}(\Sigma_{11,m}) 
	& \equiv
	\int d\Sigma_{22} ~ d\Sigma_{12} ~ P(\hat{\Sigma}_{m}) \\
	& =
	\frac{\alpha_{11}^{m/2+3/2}}{\Gamma(m/2+3/2)} e^{S_{red}{(\Sigma_{11,m})}} e^{-\alpha_{11} \Sigma_{11,m}}; \\
	%%%
	p_{22,m}(\Sigma_{22,m}) 
	& \equiv
	\int d\Sigma_{11} ~ d\Sigma_{12} ~ P(\hat{\Sigma}_{m}) \\
	& =
	\frac{\alpha_{22}^{m/2+3/2}}{\Gamma(m/2+3/2)} e^{S_{red}{(\Sigma_{22,m})}} e^{-\alpha_{22} \Sigma_{22,m}}; 
	\end{split}
\label{reduced_xxyy_distributions}
\end{equation}
Where we have introduced a reduced version of the entropy eq.~\ref{generalized_entropy} that depends on one force moment tensor component
 \begin{equation}
	\begin{split}
	S_{red}{(\Sigma_{11,m})} &= (\Sigma_{11,m})^{(m / 2+1/2)}; \\
	S_{red}{(\Sigma_{22,m})} &= (\Sigma_{11,m})^{(m / 2+1/2)}.
	\end{split}
\label{reduced_entropy}
\end{equation}
% \begin{equation}
%	\begin{split}
%	P_{11,m}(\Sigma_{11}) &=\frac{\alpha_{11}^{m \ A+3/2}}{\Gamma(m \ A+3/2)} \Sigma_{11}^{m \ A+1/2} e^{-\alpha_{11} \Sigma_{11}}; \\
%	P_{22,m}(\Sigma_{22}) &=\frac{\alpha_{22}^{m \ A+3/2}}{\Gamma(m \ A+3/2)} \Sigma_{22}^{m \ A+1/2} e^{-\alpha_{22} \Sigma_{22}}.
%	\end{split}
%\end{equation}

This form of the distribution for $ \Sigma_{11,m}$ and $ \Sigma_{22,m}$ at various $\hat{\alpha}$ can be collapsed by defining the  rescaled  dimensionless variables
\begin{equation}
\begin{split}
x &= \frac{1}{m} \alpha_{11} \Sigma_{11,m} \ or \\
x &= \frac{1}{m} \alpha_{22} \Sigma_{22,m},
\end{split}
\label{rescaled_local_fmt}
\end{equation}
yielding a gamma distribution
\begin{equation}
g_m(x) = \frac{m/2+3/2}{\left(m/2\right)^{m/2+3/2}} ~ x^{m/2+1/2}~ e^{-m x}.
\label{gamma_functions}
\end{equation}

%{\color{red} REWRITEREWRITEREWRITEREWRITEREWRITE}
%fig.~\ref{combo_subregion_entropy}(b) shows that the $\Sigma_{11}$ in SJ states can indeed be collapsed by the rescaling, the same holds for the $22-$component but is not shown. 
%%%%%%%%%%
%\begin{figure}[htbp]
%\includegraphics[width=0.49\textwidth]{collapse}
% \caption{m=15 case}
% \label{collapse}
%\end{figure}
%%%%%%%%%%%%%
%where $x = \Sigma_{12,m} / \sqrt{\Sigma_{11,m} \ \Sigma_{22,m}}$.

%{\color{red} Where did the parameter $A$ come from? How is it chosen?  What values can it have?}

\begin{figure}[htbp]
\center
\includegraphics[width=0.4\textwidth]{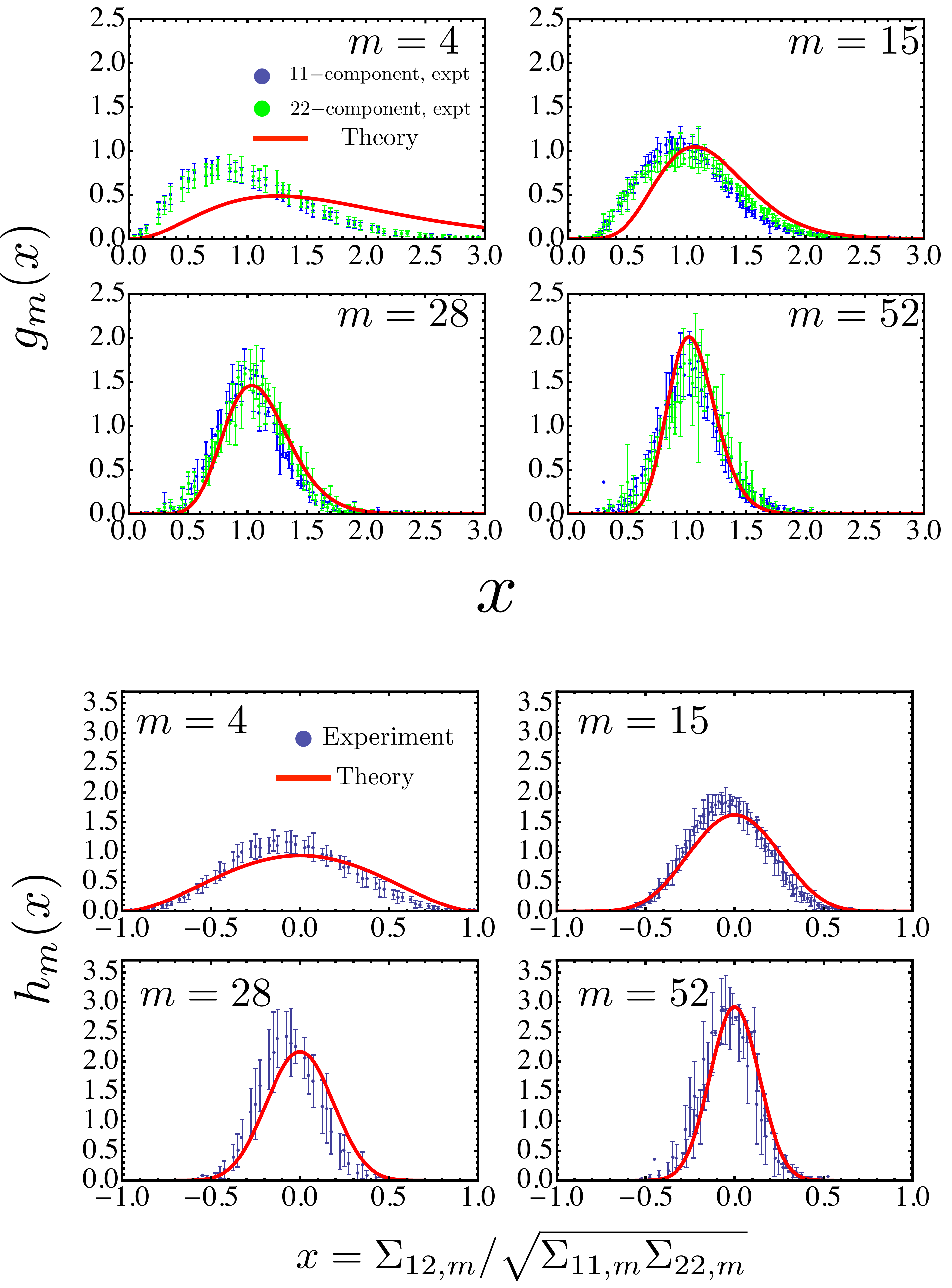}
 \caption{
 (a)
 Probability distributions for the reduced variable defined in eq.~\ref{rescaled_local_fmt}. Solid red curves are theoretical predictions (eq.~\ref{gamma_functions} with fitting parameter $A=1/2$) compared to SJ experimental data. $m$ indicates number of grains in the subregion over which the local force moment tensor is defined.
 (b)
 Probability distributions for the reduced variable defined in eq.~\ref{reduced_xy_variable}. Solid red curves are theoretical predictions  compared to SJ experimental data. $m$ indicates number of grains in the subregion over which the local force moment tensor is defined.
 }
 \label{fig:combo_reduced_disb}
\end{figure}

The rescaled distributions from SJ states are  compared to the prediction eq.~\ref{gamma_functions} in fig.~\ref{fig:combo_reduced_disb}(a) for various local subregion sizes ranging from $m=5$ to $71$. This comparison shows that the theoretical prediction captures the mean and overall shape of the experimental distributions, but that theory and experiment differ significantly in the width of the distributions.  
%\begin{figure}[htbp]
%\includegraphics[width=0.49\textwidth]{sigma_xxyy_scaling.pdf}
% \caption{}
% \label{fig:sigma_xxyy_scaling}
%\end{figure}
%% {\color{red} discussion about why the variance don't fit add figure for variance vs. m to demonstrate the idea. }
%
%\begin{figure}[htbp]
%\includegraphics[width=0.49\textwidth]{sigma_xy_scaling.pdf}
% \caption{Probability distributions for the reduced variable defined in eq.~\ref{reduced_xy_variable}. Solid red curves are theoretical predictions (eq.~\ref{xy_reduced_disbn} with fitting parameter $A=1/2$) compared to SJ experimental data. $m$ indicates number of grains in the subregion over which the local force moment tensor is defined.}
% \label{fig:sigma_xy_scaling}
%\end{figure}

There is no simple scaled form for the off-diagonal component $\Sigma_{12,m}$, but in the limit of eq.~\ref{small_xy_limit} or  $\alpha_{12} \to 0$, we can write a reduced distribution of $\Sigma_{12,m}$ in terms of rescaled local force moment components 
\begin{equation}
x = \Sigma_{12,m} / \sqrt{\Sigma_{11,m} \ \Sigma_{22,m}}
\label{reduced_xy_variable}
\end{equation} 
to obtain
\begin{equation}
h_{m}(x)=\frac{\Gamma( m \ /2 +3/2)}{\Gamma(m \ /2+1)} \left( 1-x^2 \right)^{m / 2}.
\label{xy_reduced_disbn}
\end{equation}
In fig.~\ref{fig:combo_reduced_disb}(b), we plot eq.~\ref{xy_reduced_disbn} for several $m$-values compared with SJ data. In this case, there is very good agreement between theory and experiment, including the variance.

\subsection{Discussion}

%{\color{red} explain the limit at which this is valid. also justify why disbn is reduced in this special way. and why it is applicable to SJ states, also the shift in xy fits are due to this assumption, not a problem with the model. only this assumption allows for easy comparison. } 

\begin{figure}[htbp]
\center
\includegraphics[width=0.4\textwidth]{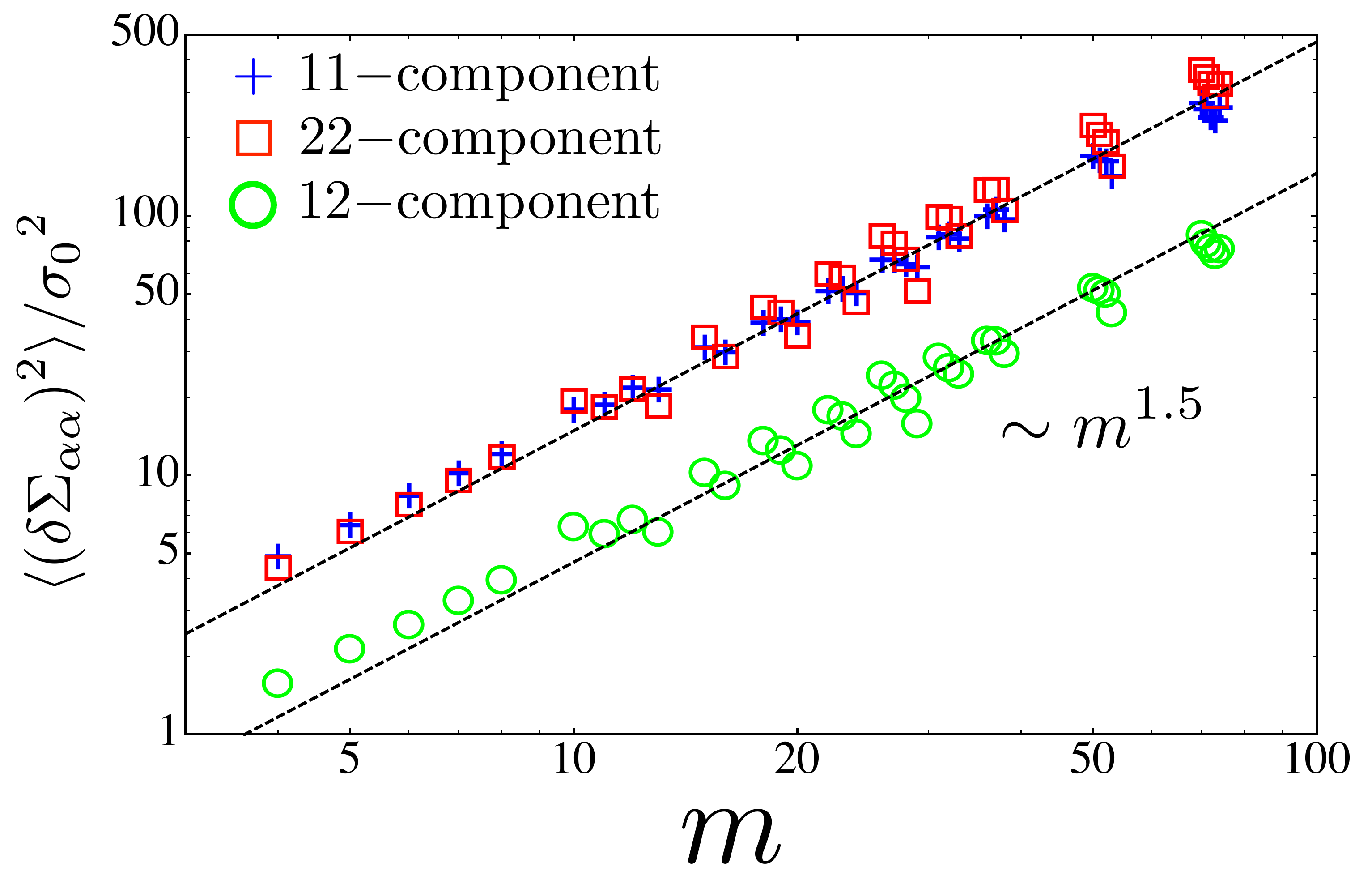}
\caption
{
$\log-\log$ plot of the variances for each local force moment tensor component as function of $m$. The variances are scaled by arbitrary stress scales to make them dimensionless. Dotted lines indicate the trend of $m^{1.5}$. 
}
\label{variance}
\end{figure}
%%%

Fig.~\ref{fig:combo_reduced_disb}(a) clearly shows that the variance of the local force moment tensor $\hat{\Sigma}$ is not well described by an ideal-gas type model,particularly for small $m$.   To explore this deviation further, we studied the scaling of the variance with the number of grains, $m$.   As shown in fig.~\ref{variance}, the variances of all components of the stress scale as $m^{\gamma}$, with $\gamma \approx 1.5$.   We note that the variance is related to the integral of the two-point, spatial correlation function of the force-moment tensor of grains.  The observed scaling, with an exponent larger than unity, can only arise if the correlations are long-ranged and do not decay within the region containing $m$ grains.  \cite{Kardar_book_2}. Since our scaling is valid for $m \approx 100$, which corresponds approximately to a $10 \times 10$ grain subregion, we can deduce that stress correlations do not decay away by $10$ neighbors.  
%The value of $\gamma$ is consistent with the variance of $x$ implied by  eq. \ref{xy_reduced_disbn}, and Fig. \ref{fig:combo_reduced_disb}(b), if we assume that fluctuations of different components of stress are uncorrelated.  
The observed  scaling of the variance of stress components in the SJ states is in sharp contrast to the experimental observation in IJ states, where the variance scales as $m$\cite{Daniels-Puckett}.  Interestingly, the variance of the scaled stress defined in  eq.~\ref{reduced_xy_variable} scales as $m$ indicating that this scaled stress does not have long-range correlations. 
For both IJ and SJ states, there is a well defined thermodynamic limit. SJ states converge with $\sqrt{\langle (\delta \Sigma)^2 \rangle} / \langle\Sigma\rangle \sim m^{-0.25}$, while IJ states converge faster with $\sqrt{\langle (\delta \Sigma)^2 \rangle} / \langle\Sigma\rangle \sim m^{-0.5}$.

Since the probability of occurrence of SJ states is well-described by the Boltzmann-like distribution of eq.~\ref{full_canonical}, we can use our experience with equilibrium statistical mechanics calculations to reconcile the adequacy of the ideal-gas like equation of state with the presence of long-range correlations.  It is known\cite{Kardar_book_2} that two-point correlations enter calculations of entropy (or free energy) through logarithmic corrections.   It is, therefore,  reasonable to expect that the entropy and the equation of state are much less sensitive to the existence of correlations than the variance of the force-moment tensor.   
%Max: I am not sure if the pressure variance is actually m^1, I think we should leave this next sentence out.
%It is also intriguing to note that the variance of the pressure in SJ states scales with $m$, and the pressure distributions agree in detail with the predictions of the ideal gas model.   
%{\color{blue} can we say the same thing for the xy component ?}  
The long-ranged correlations in the local force moment tensor  are indicative of long force chains  that are clearly visible in SJ states\cite{Nature_Bi,granmat_jie}. We are currently investigating the microscopic origin of the observed long-range correlations.  

In this work, we have demonstrated that the stress ensemble provides an excellent quantitative description of fluctuations in experimental SJ states. We show that the stress fluctuations are controlled by a single tensorial quantity-the angoricity of the system, which is a direct analog of the thermodynamic temperature.  We show that the SJ states exhibit significant correlations in local stresses in sharp contrast to IJ states. This observation reinforces the conclusion\cite{Nature_Bi} that SJ states are not merely density-driven jammed states created at a lower density.   These states are inherently different from density-driven IJ states.

\acknowledgments
We would like to thank J. Puckett, K. Daniels and R. Blumenfeld for helpful discussions and comments. 
B.C. \&  D.B. acknowledge support provided by
NSF-DMR-0905880. 
J.Z. acknowledges the support from SJTU startup fund and the award of the Chinese 1000-Plan (C) fellowship. 
R.P.B. acknowledges support provided by
NSF-DMR 09060906908,
NSF-DMR 1206351,
ARO Grant No. W911NF-11-1-0110, 
and 
NASA NNX10AU01G.

\bibliographystyle{eplbib}

\end{document}